# Photonic reinforcement learning based on optoelectronic reservoir computing

**Kazutaka Kanno[1],* and Atsushi Uchida[1]**

[1]Department of Information and Computer Sciences, Saitama University 255 Shimo-Okubo, Sakura-ku, Saitama City, Saitama 338–8570, Japan

*Corresponding author: *E-mail address*: kkanno@mail.saitama-u.ac.jp

**ABSTRACT**

Reinforcement learning has been intensively investigated and developed in artificial intelligence in the absence of training data, such as autonomous driving vehicles, robot control, internet advertising, and elastic optical networks. However, the computational cost of reinforcement learning with deep neural networks is extremely high and reducing the learning cost is a challenging issue. We propose a photonic on-line implementation of reinforcement learning using optoelectronic delay-based reservoir computing, both experimentally and numerically. In the proposed scheme, we accelerate reinforcement learning at a rate of several megahertz because there is no required learning process for the internal connection weights in reservoir computing. We perform two benchmark tasks, CartPole-v0 and MountanCar-v0 tasks, to evaluate the proposed scheme. Our results represent the first hardware implementation of reinforcement learning based on photonic reservoir computing and pave the way for fast and efficient reinforcement learning as a novel photonic accelerator.

**Introduction**

Machine learning in artificial intelligence has been the primary automation tool used in processing large amounts of data in communications and information technologies [1-4]. Reinforcement learning is a machine learning scheme involved in training an action policy to maximize the total reward in a particular situation or environment [5]. Various applications have been studied for reinforcement learning, such as autonomous driving vehicles [6], robot control [7], communication security [8], and elastic optical networks [9]. Recently, many algorithms used for reinforcement learning have been actively developed. For example, an algorithm based on a deep neural network (Agent57) has been proposed in 2020 [10]. This scheme has achieved a score that is above the human baseline on all 57 Atari 2600 games. In addition, simulated policy learning is one of the model-based reinforcement learning schemes [11]. This algorithm requires fewer training time steps. Moreover, a multi-agent reinforcement learning scheme (AlphaStar) has been proposed [12]. This scheme achieves real-time processing at 30 ms and almost outperforms human players in the online game *StarCraft II*.

Deep neural networks have often been used for reinforcement learning based on $Q$-learning, known as the deep $Q$ network [13]. The deep $Q$ network is trained to produce the value of action in a particular state. The technique of deep $Q$ networks has contributed to the development of reinforcement learning. However, learning the connection weights of deep neural networks using reinforcement learning entails high computation costs because of the repeated training of network weights from vast playing data [14,15]. This fact indicates the need for a large number of parameters used for learning to improve the performance of deep neural networks, known as overparameterization [15-17]. Large-scale overparameterization has several hundred million parameters for deep learning [15], and the training time is required for days or weeks using the deep $Q$ network on GPU [15]. Several techniques have been proposed to reduce learning costs, such as prioritized experienced replay [18]. However, the prioritized experienced replay speeds up only by a factor of two. A more efficient implementation than deep neural networks is required for reinforcement learning.

Reservoir computing has attracted significant attention in various research fields because it is capable of fast learning that results in reduced computational/training costs compared to other recurrent neural networks [19,20]. Reservoir computing is a computation framework used for information processing and supervised learning [21,22]. The main advantage of reservoir computing is that only the output weights (readout weights) are trained using a simple learning rule, realizing a fast-learning process, and enabling a reduction in its computational cost.

Recently, physical implementations of reservoir computing and its hardware implementations have been intensively studied [23-28]. Specifically, the photonic implementation of reservoir computing based on the idea of photonic accelerators [29] can realize fast information processing with low learning costs [30-35]. A previous study reported the realization of speech recognition at 1.1 GB/s using photonic reservoir computing [36]. This result suggested the reduction of computational cost and fast processing speed in photonic reservoir computing. However, photonic reservoir computing has been applied to supervised learning, and no hardware implementation of reservoir computing for reinforcement learning has been reported yet.

Hardware implementations including photonics for reinforcement learning have a demand for the applications in edge computing. In edge computing, data processing is executed close to the data source without connecting to a powerful server computer through a network [37]. Edge computing requires low power, memory budget, processing speed, and efficiency [37]. Therefore, the hardware implementation of reinforcement learning based on photonic reservoir computing is a promising candidate for edge computing.

Here, we demonstrate the photonic on-line implementation of reinforcement learning based on optoelectronic delay-based reservoir computing, both experimentally and numerically. The photonic reservoir computing is implemented based on an optoelectronic time-delayed system [30,38,39] and is used to select an agent's action to evaluate the action-value function. The output weights in reservoir computing are trained based on the reward obtained from the reinforcement learning environment, where $Q$-learning is used to update the output weights in reservoir computing. We perform two benchmark tasks, CartPole-v0 and MountainCar-v0, for the evaluation of our proposed scheme. Our demonstration is a novel on-line hardware implementation of reinforcement learning based on photonic reservoir computing.

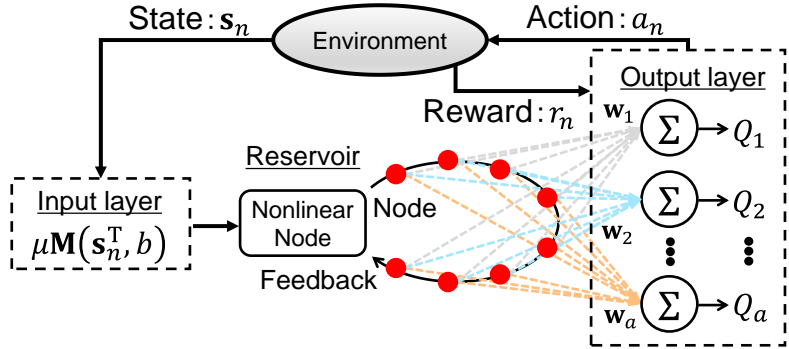

**Figure 1** Schematic diagram of reinforcement learning based on delay-based reservoir computing.

**Results**

**Reinforcement learning based on reservoir computing**

Figure 1 shows a schematic of reinforcement learning based on reservoir computing, incorporating a decision-making agent and an environment [5]. The agent affects the future state of the environment by its actions, and the environment provides rewards to every action of the agent. The objective of the agent is to maximize the total reward. However, the agent has no information regarding a good action policy. Here, we consider the action-value function $Q(\mathbf{s}_n, a_n)$ for state $\mathbf{s}_n$ and action $a_n$ at the $n$-th time step [5]. The agent selects an action with the highest $Q$ value in each state, and the total reward is increased if the agent initially knows the value of $Q$. However, the $Q$ function is usually unknown. In various previous studies, the $Q$ function was replaced by deep neural networks. Deep neural networks were trained to approximate the $Q$ function using some methods, including $Q$-learning [13]. Here, the $Q$ function is replaced by photonic delay-based reservoir computing to reduce the learning cost and realize fast processing. The reservoir computing consists of three layers: input, reservoir, and output. We explain about the three layers for reinforcement learning. Table 1 summarizes the variables used in the following explanation.

Table 1. Parameters used for reinforcement learning.

| Symbol | Parameter |
|---|---|
| $N$ | Number of virtual nodes in reservoir |
| $\tau$ | Delay time of reservoir |
| $\theta$ | Node interval used for time-multiplexing |
| $\mathbf{s}_n$ | Vector for environmental state at the time step $n$ for reinforcement learning task |
| $s_{i,n}$ | $i$-th element of the state vector $\mathbf{s}_n$ |
| $N_s$ | Number of elements for state vector $\mathbf{s}_n$ |
| $\mathbf{M}$ | $N \times (N_s + 1)$ matrix for input mask in preprocessing input signal |
| $m_{p,q}$ | Element of the mask matrix $\mathbf{M}$ in row $p$ and column $q$ |
| $\mathbf{u}_n$ | Input vector generated from the state $\mathbf{s}_n$ after preprocessing |
| $u_{i,n}$ | $i$-th element of the input vector $\mathbf{u}_n$ |
| $\mu$ | Input scaling coefficient introduced in preprocessing |
| $b$ | Input bias introduced in preprocessing |
| $u(t)$ | Input signal generated from temporally stretching input vector $\mathbf{u}_n$ |
| $T_m$ | Mask period given by $N_s \times \theta$ |
| $Q(\mathbf{s}_n, a)$ | Action-value function for the state $\mathbf{s}_n$ and the action $a$ |
| $\mathbf{w}_a$ | Weight vector for the reservoir output corresponding to the action $a$ |
| $w_{i,a}$ | $j$-th element of the output weight $\mathbf{w}_a$ corresponding to the action $a$ |
| $\mathbf{v}_n$ | Node vector extracted from the temporal response of the reservoir to the input state |
| $v_{i,n}$ | $j$-th element of the node vector $\mathbf{v}_n$ for the $n$-th input state |
| $\alpha$ | Constant step-size parameter for updating the reservoir weights in $Q$-learning |
| $\gamma$ | Discount rate for future expected reward in $Q$-learning |

In delay-based reservoir computing, the reservoir consists of a nonlinear element and a feedback loop [40]. In this scheme, the nodes in the network are virtually implemented by dividing the temporal output into short node intervals $\theta$. The number of nodes $N$ is given by $N = \tau/\theta$, where $\tau$ is the feedback delay time of the reservoir. The definition of the virtual nodes results in an easier implementation because it does not require preparing many spatial nodes to construct a network.

In the input layer, the $n$-th input data into the reservoir is the state vector given by the environment $\mathbf{s}_n^T = (s_{1,n}, s_{2,n}, \cdots, s_{N_s,n})$, where $N_s$ is the number of state elements and the superscript $T$ represents the transpose operation. The state vector is preprocessed via the masking procedure before injecting into the reservoir as follows [40,41]:

$$\mathbf{u}_n^T = (\mu s_{1,n}, \mu s_{2,n}, \ldots, \mu s_{N_s,n}, b)\mathbf{M} = (\mu \mathbf{s}_n^T, b)\mathbf{M}, \tag{1}$$

where $\mathbf{M}$ is the mask matrix with $N \times (N_s + 1)$ elements, $\mu$ is the scaling factor for the input state $\mathbf{s}_n$. The value of the mask is randomly obtained from a uniform distribution of $[-1, 1]$. The mask acts as random connection weights from input data to reservoir nodes. We represent the $i$-th element of the preprocessed input vector $\mathbf{u}_n$ as $u_{i,n}$. The vector element $u_{i,n}$ corresponds to the input data into the $i$-th virtual node. An input signal injected into the reservoir is generated by temporally stretching the elements of $\mathbf{u}_n$ to the node interval $\theta$ as follows:

$$u(t + (n-1)T_m) = u_{i,n} \ ((i-1)\theta \le t < i\theta), \tag{2}$$

where $T_m$ is the signal period of each input data and is called the mask period. The period $T_m$ is given by $T_m = N\theta$ and corresponds to the feedback delay time of the reservoir $\tau$. The input signal $u(t)$ is injected into the reservoir to generate a response signal.

We note that an input bias $b$ is added to Eq. (1). The input bias prevents the signal $\mathbf{u}_n$ from being equal to zero when the elements of $\mathbf{s}_n$ are close to zero. Moreover, the input bias leads to different nonlinearities for each virtual node. We consider the input data $u_{i,n}$ for the $i$-th virtual node defined as $\mu(m_{1,i}s_{1,n} + m_{2,i}s_{2,n} + \cdots + m_{N_s,i}s_{N_s,n}) + bm_{N_s+1,i}$, where $m_{p,q}$ is an element of the mask matrix $\mathbf{M}$ in the row $p$ and column $q$. The representation of $u_{i,n}$ indicates that the input data for the $i$-th node oscillates with the center on the bias $bm_{N_s+1,i}$. The center point of the oscillation in the input data is different for each node because the element $m_{N_s+1,i}$ of the mask matrix is different for each node. A different part of the nonlinear function that represent the relationship of the input and output in the reservoir is used for each node because of the bias $bm_{N_s+1,i}$, leading to different nonlinearities for each node. Therefore, adding an input bias enhances the approximation of the reservoir.

In the output layer, the output of reservoir computing is calculated from the weighted linear combination of virtual node states. The reservoir output is considered as the action-value function $Q(\mathbf{s}_n, a)$ for reinforcement learning. Then, the action-value function $Q(\mathbf{s}_n, a)$ is given as:

$$Q(\mathbf{s}_n, a) = \sum_{j=1}^{N} w_{j,a} v_{j,n} = \mathbf{w}_a^T \mathbf{v}_n, \tag{3}$$

where $v_{j,n}$ is the $j$-th virtual node state for the $n$-th input and $w_{j,a}$ is the output weight corresponding to the $j$-th virtual node for the action $a$. The vector $\mathbf{v}_n$ and $\mathbf{w}_a$ are given as $\mathbf{v}_n^T = (v_{1,n}, v_{2,n}, \ldots, v_{N,n})$ and $\mathbf{w}_a^T = (w_{1,a}, w_{2,a}, \ldots, w_{N,a})$, respectively. The number of reservoir outputs corresponds to the number of actions. In reinforcement learning, the action with the highest $Q$ value is selected.

Here, we use $Q$-learning algorithm to train the reservoir weights [5]. The update rule based on $Q$-learning is off-policy learning [5], where the action used for training differs from the selected action. In the $Q$-learning method, the maximum of the $Q$ function $\max_a Q(\mathbf{s}_{n+1}, a)$ for the action $a$ at the next state $\mathbf{s}_{n+1}$ is used, and the actual action is not always used for training. In our scheme, $Q(\mathbf{s}_n, a)$ is approximated using reservoir computing by considering a one-step temporal difference error $\delta_n = r_{n+1} + \gamma \max_a Q(\mathbf{s}_{n+1}, a) - Q(\mathbf{s}_n, a_n)$ and the square of the temporal difference error as the loss function. Then, the update rule for the reservoir weights is:

$$\mathbf{w}_{a_n} \leftarrow \mathbf{w}_{a_n} + \alpha \left[ r_{n+1} + \gamma \max_a \mathbf{w}_a^T \mathbf{v}_{n+1} - \mathbf{w}_{a_n}^T \mathbf{v}_n \right] \mathbf{v}_n, \tag{4}$$

where $\alpha$ is the constant step-size parameter and $\gamma$ is the discount rate for a future expected reward. These hyperparameters should be appropriately selected for a successful computation. We set $\alpha$ as a small positive value and is related to the training speed. Moreover, $\gamma$ is set to a positive value of less than one. The details of the training algorithm are described in the Methods section.

Our scheme is demonstrated in both numerical simulation and experiment using an optoelectronic delay system [42]. Figure 2(a) shows the schematic model of an optoelectronic delay system. The system has been applied to explore complex phenomena such as dynamical bifurcation, chaos, and chimera states [43]. Moreover, the application of this system in physical reservoir computing has also been studied [37,38]. The system is composed of a laser diode (LD), a Mach-Zehnder modulator (MZM), and an optical fiber for delayed feedback. In particular, the modulator provides a nonlinear transfer function $\cos^2(\cdot)$ from the electrical inputs to the optical outputs. The optical signal is transmitted through the optical fiber with a delay time of $\tau$ and is transformed into an electric signal using a photodetector (PD). The electric signal is fed back to the MZM after the signal passes through an electric amplifier (AMP). An input signal for reservoir computing is injected into the reservoir by coupling with the feedback signal. The temporal dynamics of the system is described using simple delay differential equations [44]. We use delay differential equations for the numerical verifications of the proposed scheme. The delay differential equations are described in the Methods section. In our experiment, we employ a system similar to the scheme shown in Fig. 2(a), except for the absence of the delayed feedback, as shown in Fig. 2(b). Thus, the proposed system is considered as an extreme learning machine, which has been studied as a machine-learning scheme [45]. The details of the experimental setup and online procedure for reinforcement learning are described in the Methods section.

In both numerical simulation and experiment, the number of nodes $N$ is 600, and the node interval $\theta$ is 0.4 ns. Then, the mask interval $T_m$ is given as $T_m = N\theta = 240$ ns. The feedback delay time is fixed at the same value as the mask interval in various studies on delay-based reservoir computing [36,40]. However, it has been reported that the slight mismatch between the delay time and the mask interval enhances the performance of information processing [30,46]. Therefore, we set the feedback delay time to $\tau = 236.6$ ns ($\tau = T_m - \theta$).

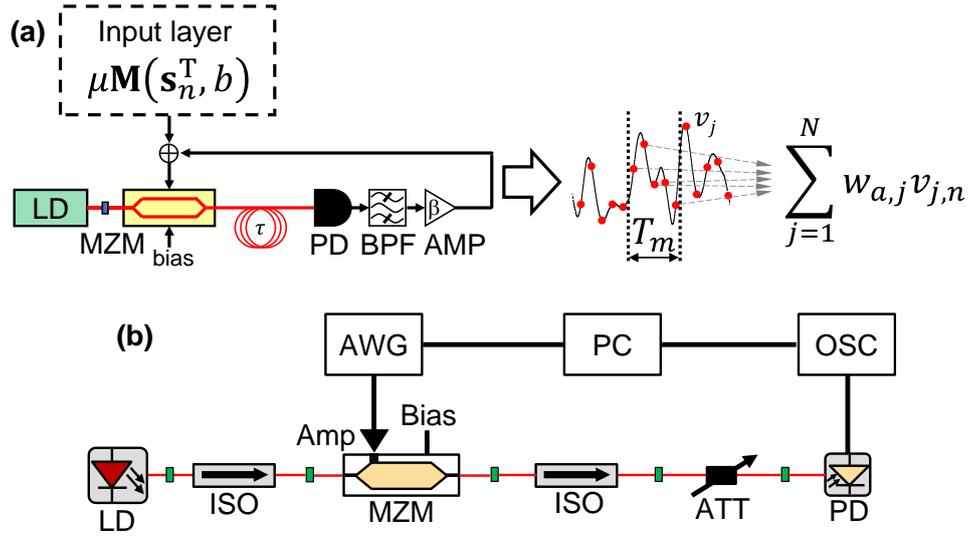

**Figure 2** (a) Schematic diagram of the optoelectronic delay system for reservoir computing. An input signal is preprocessed before injecting into the reservoir and added to a feedback signal. Reservoir node states are extracted from the temporal response of the reservoir and are shown as red circles. In the schematic diagram, MZM is the Mach-Zehnder modulator, PD is the photodetector, BPF is the bandpass filter, and AMP is the electric amplifier. (b) Experimental setup for reinforcement learning. The system has no delayed feedback, and the detected signal at the PD is not fed back to the MZM. In the personal computer (PC), environmental states in reinforcement learning tasks are calculated and the masking procedure is applied. The input data preprocessed in the PC is transferred to the arbitrary waveform generator (AWG). The optical signal from the MZM is detected at the PD, and the detected power is adjusted using the optical attenuator (ATT). The detected signal at the PD is measured at the digital oscilloscope (OSC). The AWG and the OSC are controlled by the PC in an on-line manner.

**Numerical and experimental results of reinforcement learning using benchmark tasks**

We evaluate our reinforcement learning scheme based on delay-based reservoir computing using a reinforcement learning task, known as CartPole-v0 in OpenAI Gym [47]. An un-actuated joint attaches a pole to a cart that moves along a frictionless track. The goal of the task is to keep the pole upright during an episode. An episode has a length of 200 time steps. A reward of $+1$ is provided for every time steps if the pole remains upright. The task is solved when the pole remained upright for 100 consecutive episodes. The details of the CartPole-v0 task are described in the Methods section.

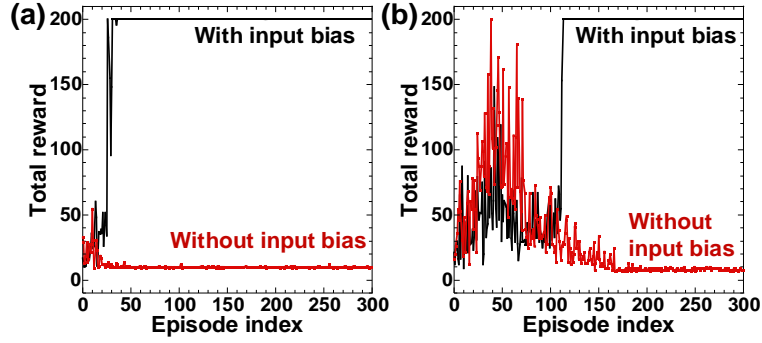

**Figure 3** (a) Numerical and (b) experimental results of the CartPole-v0 task. The result shows the total reward for each episode. The total reward of 200 indicates that the pole keeps upright over an episode. The black and red curves show the case with and without the input bias ($b = 0.8$ and $b = 0.0$), respectively.

Figure 3(a) shows the numerical results of the total reward as the episode is increased for the CartPole-v0 task. The total reward of 200 indicates that the pole remains upright over an episode and the task is solved successfully. We compare the cases with and without the input bias $b$. The input bias is applied ($b = 0.80$) for the case of the black curve in Fig. 3(a). The pole cannot be kept upright for the first several episodes. However, the total reward becomes 200 and the pole becomes upright as the number of episodes increases. The total reward of 200 is always obtained in consecutive 100 episodes from the 30th to 300th episode. Therefore, the CartPole-v0 task is successfully solved. However, for the case without input bias ($b = 0$, the red curve), the total reward does not reach 200 and the pole cannot keep upright for all episodes. The comparison of the black and red curves indicates that the input bias is required to solve the task. When no input bias is introduced, it was observed that only one action (push to the left or right) is selected regardless of the state. When the input bias is introduced, the action that prevents the pole from tilting is selected. It is considered that the input bias contributes to training the reservoir so that the reservoir can identify the state.

Our scheme requires 130 episodes for solving the task and it is faster than the result presented in [48], where more than 150 episodes are required for solving the task using a deep neural network with prioritized experienced replay. Another scheme using double deep neural network provided a similar performance as our scheme [49]. The double deep neural network requires a similar number of episodes to our scheme; however, the number of training parameters is large (150,531 parameters). Therefore, our scheme requires less training costs than these existing schemes.

Figure 3(b) shows the experimental result for the CartPole-v0 task. The input bias is introduced for the black curve. The total reward reaches 200 at the 110th episode and keeps the total reward until the 300th episode, indicating that the task is successfully solved in the experiment. We found that the total reward increases slowly in the experimental result than in the numerical result. We speculate that the measurement noise in the experiment perturbs the $Q$ value estimated by the reservoir. The noise prevents the increase of the total reward. A proper action may not be selected due to the influence of noise when the difference between the $Q$ values of the two actions is too small. Therefore, it is necessary to learn the $Q$ values until their difference becomes sufficiently large to ensure the selection of a proper action in a noisy environment. In addition, time-delayed feedback may affect the speed of increase in the total reward. The time-delayed feedback provides a memory effect for the reservoir. If the reservoir has a memory effect, it can learn the state-action value function including the past states. The introduction of time-delayed feedback is equivalent to expanding the dimension of the input state space and can approximate a complex state-action value function. Here, no time-delayed feedback is introduced in the experiment, therefore, the speed of increase in the total reward in the experiment is slower than that in the numerical simulation.

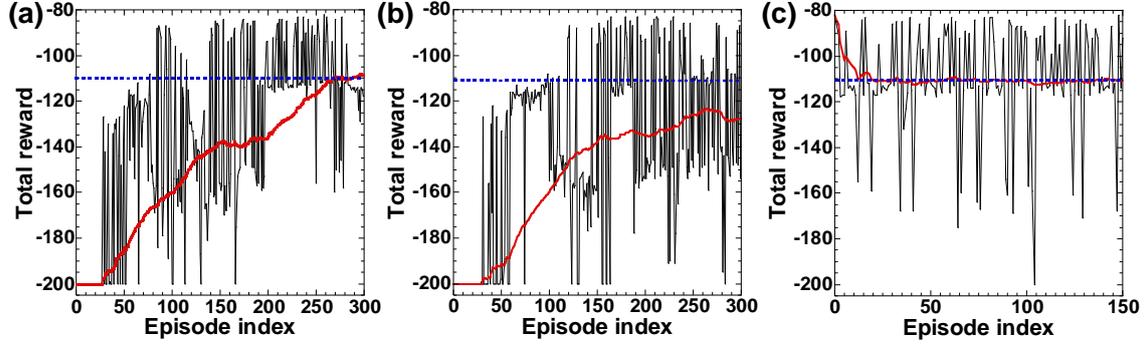

**Figure 4** (a) Numerical and (b) experimental results of the MountainCar-v0 task. The black curve represents the total reward for each episode. The moving average of the total reward is represented as the red curve. The average is calculated from the past 100 episodes. The total reward of $-200$ indicates that the car does not reach the top of the mountain. A larger value of the total reward indicates that the car reaches the top of the mountain at a smaller number of steps. (c) The total reward for each episode, where the reservoir weight at the $180^{th}$ episode in (b) is used in experiment. The weight is not updated in (c). The blue dashed line corresponds to the total reward of $-110$ that indicates that the task is successfully solved.

We demonstrate another benchmark task called the MountainCar-v0 task provided by OpenAI Gym [47]. This task aims to make a car reach the top of the mountain by accelerating the car to the right or left. One episode for the task consists of 200 steps. A reward of -1 is given for every step until an episode ends. An episode is terminated if the cart reaches the top of the mountain. Therefore, a higher value of the total reward is obtained if the cart reaches the top of the mountain faster. Solving this task is defined as obtaining an average reward of -110.0 for 100 consecutive trials [47]. The details of MountainCar-v0 task are described in the Methods section.

Figure 4(a) shows the numerical results of the total reward as the episode is increased for the MountainCar-v0 task. The black curve in Fig. 4(a) shows the total reward for each episode, and the red curve represents the moving average of the total reward calculated from the past 100 episodes. The total reward is -200 in the first several episodes, indicating that the car does not reach the top of the mountain at all. The total reward increases as the number of episodes increases, indicating the car reaches the top of the mountain. The average reward exceeds -110 at the 267th episode, indicating that the task is solved using our scheme.

Figure 4(b) shows the experimental result for the MountainCar-v0 task. The moving average of the total reward (red curve) increases as the number of episodes increases. However, the moving average does not reach the blue dashed line (the total reward of -110). The number of consecutive episodes at which a high value of the total reward is obtained is small in the experiment. For example, a large value of the total reward from -120 to -80 is obtained during 23 consecutive episodes from the 170th episode in the black curve of Fig. 4(b). However, the moving average (red curve) cannot reach the reward of -110. The reason the reservoir cannot obtain a large value of the total reward is due to a negative value of the reward.

The negative value of the reward makes the reservoir trained not to select an action in a state. Therefore, as the episodes continue at which a large value of the total reward is obtained, the action policy for getting a large value of the total reward becomes not to be selected.

We consider utilizing a fixed reservoir weight to prevent from decreasing the total reward due to a negative reward. We use the reservoir weights obtained at the 180th episode in the experiment of Fig. 4(b), and the weights are fixed during the experiment, i.e., the weights are not updated in the training procedure. Figure 4(c) shows the total reward for each episode in this experiment. The moving average of the total reward (red curve) exceeds -110 at the 141st episode. Therefore, the task is solved if the weights are not updated. Additionally, this indicates that the trained weight works appropriately if the experimental setup conditions are slightly changed, such as the detected power at the PD. Therefore, the trained weights are robust against perturbations in the system parameters.

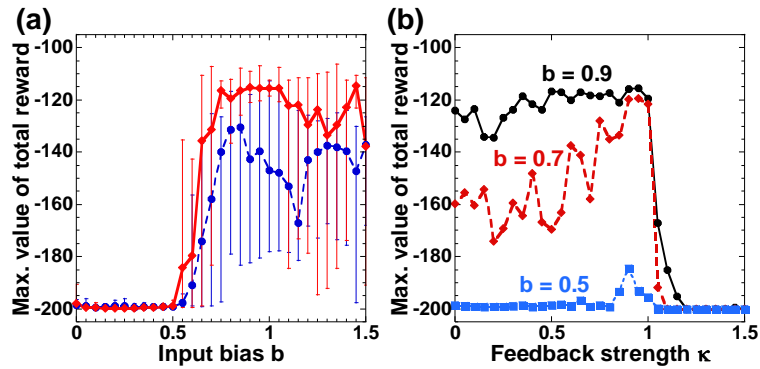

**Figure 5 (a)** Maximum of the average total reward as a function of the input bias $b$. The red solid (with diamonds) and blue dashed (circles) curves represent the case with ($\kappa = 0.9$) and without ($\kappa = 0$) delayed feedback. The average total reward is calculated using a moving window from the past 100 episodes. The error bar represents the maximum and minimum values for 10 trials. (b) Maximum of the average total reward as a function of the feedback strength $\kappa$. The input bias is set to be $b = 0.9, 0.7,$ and $0.5$ for the black solid (circles), red dashed (diamonds), and blue dotted (squares) curves, respectively. The plotted value is the maximum of the average total reward in 100 consecutive episodes. The error bar represents the maximum and minimum values for 10 trials.

We numerically investigate the dependence of the performance on the input bias in the MountainCar-v0 task. Figure 5 shows the numerical results of the maximum value of the average total reward as the input bias $b$ is changed for the MountainCar-v0 task. In Fig. 5(a), the solid red curve represents the maximum moving average of the total reward in 1000 episodes. The maximum total reward is averaged for ten trials, with each trial consisting of 1000 episodes. The total reward is close to zero for a small input bias ($b \leq 0.5$). A large total reward value is obtained for a large input bias ($b > 0.5$). This result indicates that the input bias is necessary for solving the task. The input bias with a value close to one is suitable for increasing the total reward. The result is related to the normalized half-wave voltage ($V_\pi$) of the MZM, where the normalized voltage is equal to one in our numerical simulation. The input bias nearly equal to one can produce the nonlinearity of the MZM ($\cos^2(\cdot)$), and the nonlinearity can assist in identifying different input states.

Furthermore, we investigate the effect of the time-delayed feedback in the numerical simulation. In Fig. 5(a), the blue dashed curve represents the maximum moving average of the total reward generated using the reservoir without time-delayed feedback ($\kappa = 0$). At the input bias of $b = 0.85$, the total reward of -130.29 is obtained. Thus, the reservoir successfully trains for the car to reach the top of the mountain if the reservoir has no delayed feedback. However, the performance is lower than in the case with delayed feedback (the solid red curve). Therefore, the presence of the time-delayed feedback can enhance the performance of reinforcement learning.

For more detailed investigation, Figure 5(b) shows the dependence of the total reward on the feedback strength $\kappa$. Different values of the input bias are used for each of the three curves. For the small value of the input bias (blue curve, $b = 0.50$), the total reward is almost equal to -200 for different feedback strengths. This result indicates that the adjustment of the feedback strength cannot enhance the performance when the input bias is too small. For the black ($b = 0.90$) and red ($b = 0.70$) curves, a large value of the total reward is obtained at the feedback strength of approximately one. However, the total reward decreases as the feedback strength increases over one. When the feedback strength becomes larger than one, the temporal dynamics of the optoelectronic system changes from a steady state to a periodic oscillation, though the reservoir has no input signal. The reservoir may produce different response signals to the same driving inputs when the temporal dynamics of the reservoir is periodic or chaotic. Therefore, the reservoir does not have consistency [50], which is the reproducibility of response signals in a nonlinear dynamical system driven repeatedly by a same input signal. If there is no consistency in the response signals of the reservoir, the reservoir cannot successfully learn the $Q$ function since different input states cannot be identified. Therefore, the reservoir provides high performance at the vicinity of the bifurcation point $\kappa = 1$, called the edge of chaos. In reservoir computing, it has been reported that the condition of the edge of chaos can enhance the performance in many studies [51]. In addition, our results show that the performance for reinforcement learning can be enhanced at the edge of chaos.

**Discussion**

We introduced an input bias for preprocessing the input state in reinforcement learning. The input bias has the same role as a bias introduced in the general neuron models that controls the firing frequency. Our results show that the input bias is necessary for solving the reinforcement tasks in our scheme. Here, the activation function of the virtual node of the reservoir is $\cos^2(\cdot)$, and an input bias is used to control the initial position of $\cos^2(\cdot)$ function. For example, if the input bias is set near the extreme value of $\cos^2 \psi$ ($\psi = 0, \pm n\pi/2$, and $n$ is an integer), the reservoir does not respond well with respect to the change in the input signal. In contrast, when the input bias is set to a quadrature point ($\psi = \pm \pi/4$), the reservoir shows a large response with respect to the change in the input signal. Therefore, it is possible to adjust the sensitivity of the virtual nodes to the input signal by changing the input bias. In the presence of the input bias, different input states are distinguished well, which enhances the performance of reinforcement learning based on reservoir computing. We show that input bias has a significant effect on reinforcement learning in our scheme.

We emphasize that one action of reinforcement learning is potentially determined by the processing rate of reservoir computing at a frequency of 4.2 MHz in our scheme, where one virtual network is constructed from a time series with $N\theta = 240$ ns ($N$ is 600 and $\theta$ is 0.4 ns). Further, we increase the processing speed by decreasing the node interval $\theta$ with a faster photonic dynamical system. In addition, the size of the trained parameters (600) is smaller than that for deep neural networks (e.g., 480 Mega parameters for ImageNet [15-17]). The hardware implementation of photonic reservoir computing is promising for realizing fast and efficient reinforcement learning.

The number of training parameters is reduced in reinforcement learning based on reservoir computing, compared with deep neural networks. However, reservoir computing may produce less performance than deep neural networks for more complex tasks. Therefore, our future works are the application of our scheme to more complex tasks and the comparison with conventional algorithms based on deep neural networks. In addition, the effect of memory provided by the reservoir is an important issue. Memory capacity is one of the essential characteristics of reservoir computing. The reservoir that incorporates past information to train the Q function could perform better on the tasks that require long-term memory. Therefore, the investigation of the memory effect of the reservoir on the performance of reinforcement learning is another research topic in future work.

To summarize our study, we numerically and experimentally demonstrated the on-line implementation of reinforcement learning based on optoelectronic reservoir computing, which consists of a laser diode, a Mach-Zehnder modulator, and a fiber delay line. We demonstrated two benchmark tasks of the CartPole-v0 and MoutainCar-v0 tasks using our proposed scheme. The results show that the state-action value function in reinforcement learning is trained, and their tasks are solved successfully using photonic reservoir computing. To the best of our knowledge, this is the first on-line hardware implementation of reinforcement learning based on photonic reservoir computing. In particular, reservoir computing is used to approximate the Q-function, and the output weights of the reservoir are trained with Q-learning. The high-dimensional mapping between the states and Q-values for reinforcement learning is trained by reservoir computing. The speed of one action is determined by the processing rate of reservoir computing at 4.2 MHz (240 ns) in our experiment.

The hardware implementation of reinforcement learning based on photonic reservoir computing is promising for fast and efficient reinforcement learning as a novel photonic accelerator. Our scheme can be applied for edge computing in real-time distributed control and adaptive channel selection in optical communications.

**Methods**

**Details of training algorithm for reinforcement learning**

We present a training procedure for reinforcement learning in this section. We consider that a state in a reinforcement learning task is updated at every step, and the step index is $n$. The update is repeated until termination conditions for the task are satisfied. One episode consists of all steps until the task is completed. In the algorithm, the reservoir weight $\mathbf{w}_a$ is initialized with a value randomly sampled from a uniform distribution of [-0.1 0.1]. In each episode, the following procedure is repeated from the step index $n = 1$ until termination conditions are satisfied. The state of the task is initialized, which is regarded as $\mathbf{s}_1$. The input signal $u(t)$ injected into the reservoir is generated by preprocessing the state $\mathbf{s}_n$ using Eqs. (1) and (2). The input signal $u(t)$ is injected into the reservoir and the response signal of the reservoir is obtained. A node state $\mathbf{v}_n$ is extracted from the response signal. The $Q$ value corresponding to each action $a$ is calculated from the node state $\mathbf{v}_n$ and the reservoir weight $\mathbf{w}_a$ using Eq. (3). The action $a$ with the highest $Q$ value is selected at the step index $n$. The state in the task is updated using the selected action $a_n$. Then, a reward $r_{n+1}$ and the next state $\mathbf{s}_{n+1}$ is given. A set of the states, action, and reward $(\mathbf{s}_n, \mathbf{s}_{n+1}, a_n, r_{n+1})$ is preserved. The reservoir weight is updated using Eq. (4). The step index is updated from $n$ to $n + 1$. The above procedure is repeated until termination conditions are satisfied. The total reward is given from the sum of the rewards in all steps. Algorithm 1 shows the pseudo code corresponding to the above procedure.

In the training process, we employ the experienced replay method [52]. In this method, the observed data (state, action, and reward) are preserved in the memory, sampled randomly, and used for training. The randomly sampled data is referred to as the mini-batch. The size of the mini-batch and the number of preserved data are hyperparameters. Using the randomly sampled and preserved data for training may reduce the correlation between the data used for training and exhibits easier convergence of the $Q$-learning. The number of memories and the size of the mini-batch for experience replay are 4,000 and 256, respectively.

Moreover, we used the $\varepsilon$-greedy method for the action selection. The value of $\varepsilon$ is reduced as the number of episodes increases. The value of $\varepsilon$ is updated by $\varepsilon = \varepsilon_0 + (1 - \varepsilon_0)\exp(-k_\varepsilon n_{ep})$, where $n_{ep}$ is the episode index of the reinforcement learning task and $k_\varepsilon$ is the attenuation coefficient. The coefficient $k_\varepsilon$ is fixed at $k_\varepsilon = 0.04$ here. The value of $\varepsilon$ converges to the value $\varepsilon_0 = 0.01$ as the number of episodes increases.

---

Algorithm 1 Pseudo code for reinforcement learning based on photonic reservoir computing.

---

Initialize a reservoir weight $\mathbf{w}_a$ with a value randomly sampled from a uniform distribution of [-0.1 0.1]
FOR $i = 1$ to the number of episodes
    Initialize a state in a reinforcement learning task
    Set a total reward to 0
    Initialize a step index $j = 1$ for the task
    REPEAT
        Get the state $\mathbf{s}_j$ in the task
        Generate an input signal $u(t)$ from preprocessing the state $\mathbf{s}_j$ using Eqs. (1) and (2)
        Input $u(t)$ into the reservoir and extracted a node state $\mathbf{v}_j$ from the response output of the reservoir
        Calculate $Q$ value for each action from the node state $\mathbf{v}_j$ and the reservoir weight $\mathbf{w}_a$ using Eq. (3)
        Select the action $a_j$ with the highest $Q$ value
        Update the state $\mathbf{s}_j$ to $\mathbf{s}_{j+1}$ using the selected action $a_j$ and getthereward $r_{j+1}$
        Preserve a set $(\mathbf{s}_j, \mathbf{s}_{j+1}, a_j, r_{j+1})$
        Update the reservoir weight from preserved sets using Eq. (4)
        Add the reward $r_j$ to the total reward
        Update $j$ to $j + 1$
    UNTIL termination conditions for the task are satisfied
    Print the total reward
END FOR

---

**Numerical model for an optoelectronic delay system**

Optoelectronic delay systems [42] have been studied for delay-based reservoir computing [30,38,39], using the following delay differential equations [43]:

$$\tau_L \frac{dx(t)}{dt} = -\left(1 + \frac{\tau_L}{\tau_H}\right)x(t) - y(t) + \beta \cos^2\left[\kappa x(t-\tau) + \frac{\pi}{4}u(t) + \phi_0\right] + \xi(t), \quad (5)$$

$$\tau_H \frac{dy(t)}{dt} = x(t), \quad (6)$$

where $x$ is the normalized output of MZM, $\tau_L$ and $\tau_H$ are the time constants describing the low-pass and high-pass filters related to the frequency bandwidths of the system components, relatively, $\beta$ is the feedback strength (dimensionless), $\phi_0$ is the offset phase of MZM, $u(t)$ is the input signal injected into the reservoir, and $\xi(t)$ is the white Gaussian noise with properties $\langle \xi(t) \rangle = 0$ and $\langle \xi(t)\xi(t_0) \rangle = \delta(t - t_0)$, where $\langle \cdot \rangle$ denotes the ensemble average and $\delta(t)$ is Dirac's delta function. Table 2 shows the parameter values used. A personal computer (DELL, CPU: Intel Core i7-7700 3.60 GHz, RAM: 16.0 GB, Windows 10) was used in numerical simulation.

**Table 2.** Parameter values used in numerical simulations.

| Symbol | Parameter | Value |
| --- | --- | --- |
| $(2\pi\tau_L)^{-1}$ | Low-pass cutoff frequency | $12.5 \times 10^9$ Hz |
| $(2\pi\tau_H)^{-1}$ | High-pass cutoff frequency | $0.625 \times 10^6$ Hz |
| $\tau$ | Feedback delay time | $239.6 \times 10^{-9}$ s |
| $\beta$ | Dimensionless gain | 1.0 |
| $\kappa$ | Dimensionless feedback strength | 0.9 |
| $\phi_0$ | Bias point for MZM | $-0.25\pi$ rad |
| $\mu$ | Scaling coefficient for input state | 0.6 |

**Experimental setup**

Figure 2(b) shows the experimental setup used in the experiments. The system has no delayed feedback for simple implementation, and the system corresponds to an extreme learning machine [47]. A distributed-feedback laser diode (LD, NTT electronics, NLK1C5GAAA) with an optical wavelength of 1,547 nm was used as the optical source. The lasing threshold of the LD was 11.6 mA, and the driving current was 30.0 mA. The optical output of the LD was injected into a Mach-Zehnder modulator (MZM, EO Space, AX-0MKS-20-PFA-PFA-LV-UL), where a bias controller (BC, IXBlue, MBC-AN-LAB) was inserted to stabilize the operation bias of the MZM. The bias was stabilized at the quadrature point. Moreover, a modulation signal was generated from an arbitrary waveform generator (AWG, Tektronix, AWG70002A, 25 GSample/s, 10 bit vertical resolution) and transferred to the MZM after amplification by an electric amplifier (AMP, IXBlue, DR-AN-10-HO). A photodetector (PD, Newport, 1554-B, 12-GHz bandwidth) was used to detect the optical signal of the MZM, and the detected power was 0.280 mW on the condition of no modulation input. The detected signal at the PD was transferred to a digital oscilloscope (OSC, Tektronix, DPO72304SX, 23 GHz bandwidth) and sampled at 50 GSample/s.

The signal amplitude injected into the MZM and the half-wave voltage $V_\pi$ of the MZM are important for successful computation. The signal amplitude is determined from the input scaling $\mu$ and the bias scaling $b$, the output amplitude of the AWG, and the amplification gain of the AMP. The output amplitude of the AWG is 0.30 V at the peak-to-peak value. The amplification gain of the AMP is typically 30 dB under small-signal conditions. The half-wave voltage of the MZM was $V_\pi = 4$ V. The input signal was preprocessed using Eq. (1) in the personal computer. The input scaling $\mu$ and the bias scaling $b$ for preprocessing is fixed at $\mu = 0.50$ and $b = 0.40$, respectively. These parameter values are different from those used for the numerical simulation because the signal amplitude in the experiments depends on these parameter values and the condition of the output amplitude of the AWG. In our experiments, the values of $\mu = 0.50$ and $b = 0.40$ produce an electric signal with an amplitude nearly equal to the half-wave voltage of the MZM. The condition of the bias scaling $b$ is consistent with the value for successful computation in our numerical simulation (see Fig. 5).

**Experimental online procedure for reinforcement learning**

In our experiment, the digital oscilloscope (OSC) and the arbitrary waveform generator (AWG) were controlled by the personal computer (PC). Initially, the state of the reinforcement learning task was calculated using the PC. Then, an input signal was generated from the state by applying a masking procedure for reservoir computing. The input signal was generated from the AWG. The signal was amplified by AMP and injected into the MZM. The optical output of the MZM was modulated based on the injected signal. The optical signal was transformed into an electric signal at the PD. The electric signal was acquired by the OSC and was then transferred to the PC. The node states of the reservoir were extracted from the signal. The output of reservoir computing was calculated from the weighted sum of the node states and corresponded to the $Q$ value for each action in a reinforcement learning task. An action was selected based on the $Q$ values, and the state of the reinforcement learning task was updated based on the selected action. In addition, the reservoir weights were updated based on $Q$-learning. The above procedure was repeated until the reinforcement learning task was terminated. This procedure for reinforcement learning was executed under the control of OSC, AWG, and PC in an on-line manner.

In the experiment, pre- and post-processing are implemented in the personal computer although the reservoir is hardware-implemented. Therefore, the processing speed of the experiment is restricted to the software processing in the personal computer. Here, the decision of an action was executed at about 0.5 s. The hardware implementation of the pre- and post-processing in photonic reservoir computing has been studied [53]. The processing speed can be accelerated by implementing the pre- and post-processing in hardware, such as field programmable gate array (FPGA).

**CartPole-v0 task**

The CartPole-v0 task is a benchmark task for reinforcement learning given by the OpenAI Gym [47]. In this task, we consider a pole attached by an un-actuated joint to a cart that moves along a frictionless track with four states: cart position, cart velocity, pole angle, and pole velocity at the tip. These states are initialized to uniform random values. The agent's action is to push the cart to the right (+1) and left (-1). The goal of the task is to keep the pole upright during an episode with a length of 200 timesteps, and the task is considered solved when the pole remains upright for 100 consecutive episodes. A reward of +1 is provided for every time steps while the pole remains upright. The episode ends when the pole is more than 12° from the vertical or the cart position moves more than 2.4 units or less than -2.4 units from the center. The magnitudes of the cart position and pole velocity were normalized to the range of [-1.0 to 1.0] before injecting into the reservoir. The hyperparameters for reinforcement learning are fixed at $\alpha = 0.000400$ and $\gamma = 0.995$, respectively.

**MountainCar-v0 task**

The MountainCar-v0 task is provided by OpenAI Gym [47]. The goal of this task is for a car to reach the top of the mountain by accelerating the car to the right or left. The observable states of the task are the cart position and cart velocity. In the initial state, the cart position is randomly set from a uniform distribution [-0.6 to -0.4], and the cart velocity is fixed at zero. The agent's action is to push the cart to the left, neutral, and push the cart to the right. A reward of -1 is given for every step until an episode ended. An episode consists of 200 steps and is terminated if the cart reaches the top of the mountain. The hyperparameters for reinforcement learning are fixed at $\alpha = 0.000010$ and $\gamma = 0.995$, respectively.

**Acknowledgments**

This study was supported in part by JSPS KAKENHI (JP19H00868 and JP20K15185), JST CREST JP-MJCR17N2, and the Telecommunications Advancement Foundation.

**Author contributions**

All authors contributed to the development and/or implementation of the idea.

K. K. performed the numerical simulations and analyzed the data. K. K. and A. U. contributed to the discussion of the results. K. K. and A. U. contributed to writing the manuscript.

**Competing interests**

The authors declare that they have no competing interests.